\documentstyle[12pt]{article}
\newcommand{\nc}{\newcommand}

\nc{\dbar}{\bar{\partial}}
\nc{\be}{\begin{equation}}
\nc{\ee}{\end{equation}}

\nc{\beq}{\begin{equation}}
\nc{\eeq}{\end{equation}}
\nc{\bea}{\begin{eqnarray}}
\nc{\eea}{\end{eqnarray}}

\catcode`\@=11

\@addtoreset{equation}{section}

\def\@normalsize{\@setsize\normalsize{15pt}\xiipt\@xiipt
\abovedisplayskip 14pt plus3pt minus3pt%
\belowdisplayskip \abovedisplayskip
\abovedisplayshortskip  \z@ plus3pt%
\belowdisplayshortskip  7pt plus3.5pt minus0pt}
\def\small{\@setsize\small{13.6pt}\xipt\@xipt
\abovedisplayskip 13pt plus3pt minus3pt%
\belowdisplayskip \abovedisplayskip
\abovedisplayshortskip  \z@ plus3pt%
\belowdisplayshortskip  7pt plus3.5pt minus0pt
\def\@listi{\parsep 4.5pt plus 2pt minus 1pt
            \itemsep \parsep
            \topsep 9pt plus 3pt minus 3pt}}

\def\underline#1{\relax\ifmmode\@@underline#1\else
        $\@@underline{\hbox{#1}}$\relax\fi}
\@twosidetrue
\relax

\catcode`@=12

\catcode`\@=11

\def\section{\@startsection{section}{1}{\z@}{3.5ex plus 1ex minus
   .2ex}{2.3ex plus .2ex}{\large\bf}}

\def\ps@headings{\def\@oddfoot{}\def\@evenfoot{}
\def\@oddhead{\hbox{}\hfill
        \makebox[.5\textwidth]{\raggedright\ignorespaces --\thepage{}--
        \hfill }}
\def\@evenhead{\@oddhead}
}

\ps@headings

\catcode`\@=12

\relax

\def\figcap{\section*{Figure Captions\markboth
        {FIGURECAPTIONS}{FIGURECAPTIONS}}\list
        {Fig. \arabic{enumi}:\hfill}{\settowidth\labelwidth{Fig. 999:}
        \leftmargin\labelwidth
        \advance\leftmargin\labelsep\usecounter{enumi}}}
 \relax
\def\tablecap{\section*{Table Captions\markboth
        {TABLECAPTIONS}{TABLECAPTIONS}}\list
        {Table \arabic{enumi}:\hfill}{\settowidth\labelwidth{Table 999:}
        \leftmargin\labelwidth
        \advance\leftmargin\labelsep\usecounter{enumi}}}
 \relax
\def\reflist{\section*{References\markboth
        {REFLIST}{REFLIST}}\list
        {[\arabic{enumi}]\hfill}{\settowidth\labelwidth{[999]}
        \leftmargin\labelwidth
        \advance\leftmargin\labelsep\usecounter{enumi}}}
 \relax

\catcode`\@=11

\def\ps@headings{\def\@oddfoot{}\def\@evenfoot{}
\def\@oddhead{\hbox{}\hfill
        \makebox[.5\textwidth]{\raggedright\ignorespaces --\thepage{}--
        \hfill }}
\def\@evenhead{\@oddhead}
}

\ps@headings

\relax

\def\firstpage#1#2#3#4#5#6{
\begin{document}

\begin{titlepage}
\nopagebreak
\title{\begin{flushright}
       \vspace*{-1.8in}
       {\normalsize SISSA-109/97/EP} 
\end{flushright}
\vfill
{\large \bf #3}}
\author{\large #4 \\ #5}
\maketitle
\vskip -7mm
\nopagebreak
\begin{abstract}
{\noindent #6}
\end{abstract}
\vfill
\footnotesize{PACS:  11.25 Hf} \\
\footnotesize{Keywords: D-branes, spin interactions, boundary states}
\thispagestyle{empty}
\end{titlepage}}
\newcommand{\dal}{\raisebox{0.085cm}
{\fbox{\rule{0cm}{0.07cm}\,}}}
\newcommand{\dt}{\partial_{\langle T\rangle}}
\newcommand{\dtbar}{\partial_{\langle\bar{T}\rangle}}
\newcommand{\al}{\alpha^{\prime}}
\newcommand{\mst}{M_{\scriptscriptstyle \!S}}
\newcommand{\mpl}{M_{\scriptscriptstyle \!P}}
\newcommand{\dv}{\int{\rm d}^4x\sqrt{g}}
\newcommand{\lv}{\left\langle}
\newcommand{\rv}{\right\rangle}
\newcommand{\ph}{\varphi}
\newcommand{\sbar}{\,\bar{\! S}}
\newcommand{\xbar}{\,\bar{\! X}}
\newcommand{\fbar}{\,\bar{\! F}}
\newcommand{\zbar}{\,\bar{\! Z}}
\newcommand{\tbar}{\bar{T}}
\newcommand{\ubar}{\bar{U}}
\newcommand{\ybar}{\bar{Y}}
\newcommand{\phb}{\bar{\varphi}}
\newcommand{\cm}{Commun.\ Math.\ Phys.~}
\newcommand{\pr}{Phys.\ Rev.\ D~}
\newcommand{\prl}{Phys.\ Rev.\ Lett.~}
\newcommand{\pl}{Phys.\ Lett.\ B~}
\newcommand{\ibar}{\bar{\imath}}
\newcommand{\jbar}{\bar{\jmath}}
\newcommand{\np}{Nucl.\ Phys.\ B~}
\newcommand{\e}{{\rm e}}
\newcommand{\gsi}{\,\raisebox{-0.13cm}{$\stackrel{\textstyle
>}{\textstyle\sim}$}\,}
\newcommand{\lsi}{\,\raisebox{-0.13cm}{$\stackrel{\textstyle
<}{\textstyle\sim}$}\,}
\date{}
\firstpage{95/XX}{3122}
{\large\sc\bf A note on supersymmetric D-brane dynamics} 
{Jose F. Morales$^{a}$, Claudio A. Scrucca$^{a,b}$  and Marco 
Serone$^{a,b}\footnote{Address after $1^{{\rm st}}$ september 1997: Dep. of 
Mathematics, University of Amsterdam, Plantage Muidergracht 24, 1018 TV
Amsterdam; e-mail: serone@wins.uva.nl}$}
{\normalsize\sl $^{a}$International School for Advanced Studies, ISAS-SISSA
\\[-3mm]
\normalsize\sl Via Beirut n. 2-4, 34013 Trieste, Italy\\[-3mm]
\normalsize\sl \\[-3mm]
\normalsize\sl 
$^{b}$Istituto Nazionale di Fisica Nucleare, sez.\ di Trieste, Italy\\[-3mm]
\normalsize\sl e-mail:morales,scrucca,serone@sissa.it \\[-3mm]}
{We study the spin dependence of D-brane dynamics in the Green-Schwarz formalism 
of boundary states. In particular we show how to interpret insertion of supercharges
on the boundary state as sources of non-universal spin effects in D-brane
potentials. In this way we find for a generic (D)p-brane, potentials going like 
$v^{4-n}/r^{7-p+n}$ corresponding to interactions between the different components of 
the D-brane supermultiplet. From the eleven dimensional point of view these
potentials arise from the exchange of field strengths corresponding 
to the graviton and the three form, coupled non-minimally to the branes.
We show how an annulus computation truncated
to its massless contribution is enough to reproduce these next-to-leading
effects, meaning in particular that the one-loop (M)atrix theory effective
action should encode all the spin dependence of low-energy supergravity interactions.}

\section{Introduction}
The D-brane description of solitons carrying Ramond-Ramond
charges \cite{pol} provided us with an explicit tool to study new phenomena
in string theory, improving drastically our current understanding of the
non-perturbative physics. In particular, the study of soliton interactions
or multi-soliton configurations, very non-trivial issues in quantum field
theory, are easily performed in the D-brane language. A simple one-loop
annulus computation, for instance, is enough to show the BPS ``no force''
condition between two parallel static D-branes \cite{pol} or to study the
semiclassical phase shift of one brane moving past an other \cite{bach}.
The solitons described by these brane configurations presents however
a peculiar property, not present in the more familiar solitons appearing
in quantum field theory. Their size, indeed, 
in the limit of small string coupling constant, becomes much smaller than the
usual soliton size, fixed basically by the scale of perturbative states,
allowing to test distances even shorter than the usual string length \cite{she,dkps}. 
The proposal of \cite{bfss} for a parton description of M-theory in a
given kinematical region as given by an
effective Super Yang-Mills $U(N)$ quantum mechanics \cite{nqm} 
is an exciting application of these ideas, encouraging to a 
deeper study of the dynamics of D-branes.      
Altough there have been several works analyzing D-brane interactions in
various configurations, most of them considered the approximation in which
a D-brane is a heavy semiclassical spinless state.

Aim of this work is to analyze some non-universal D-brane
interactions, due to spin effects, in order to understand the structure of the
next-to-leading terms of their potentials.
The interaction between two moving D-branes can be written schematically as    
\be
V(r^2=b^2+v^2 t^2) \sim J^{M}(0)\Delta_{MN}(r)J^{N}(r)
\ee          
where $J^{M}(r)$ denotes generically the eikonal aproximation of the currents
$J^{\mu}(r)$, $T^{\mu\nu}(r)$, etc., according to the possible spin of the
fields to which this current couples (vector field, graviton, etc.) and
$\Delta_{MN}(r)$ represents the ten dimensional propagator of the
corresponding particle exchanged. The sum over $M,N$ run over all the
infinite closed string states the two D-branes can exchange.

These currents in momentum space can be decomposed into two pieces :
\be
J^{M}(p,q)=J^{M}_{univ}(p)+J^{M}_{spin}(p,q)
\ee
where $p$ is the momentum of the scattered D-brane, in this aproximation 
much bigger than $q$, the transfered one. The universal current $J^{M}_{univ}$
is always determined simply by the ten-momentum $p$ as 
\bea
J^{\mu}\sim p^{\mu}\nonumber,\qquad
T^{\mu\nu}&\sim& p^{\mu}p^{\nu}\nonumber, \ \ \ {\rm etc.}
\eea
for currents coupled respectively to fields of spin 1, 2, etc..
The large-distance potential is governed by the universal couplings of
these currents with the massless string states, that combine in the leading
contribution \cite{bach}:
\be
V(r,v)\sim (\cosh 2v\mp\,4\cosh v+\,3)/r^{7-p} \label{vrv}
\ee
once one substitutes the ten dimensional momentum 
$p_{\mu}=M(\cosh v,\sinh v,0,....0)$. The minus (plus) sign is for branes
with the same (opposite) charge and leads, in the nonrelativistic limit, to a
brane-brane potential going like $v^4/r^{7-p}$
\footnote{Being interested in nonrelativistic processes, we do not distinguish 
between velocity and rapidity throughout all the paper.}.
This is indeed the universal, spin-independent potential for two moving 
D-branes.

The remaining part of the current $J_{M}^{spin}$,
at least linear in the transfer momentum, will lead to subleading 
potentials at large distances whose specific form will depend on the
particular state of the D-brane supermultiplet.
These sources represent in general non-minimal couplings of super D-branes
to the bulk closed string states. In this note we study these
next-to-leading effects by using the boundary state technique in the
Green-Schwarz formalism \cite{gregut1}. We introduce
our formalism, showing how to define a moving boundary state in light-cone
gauge and the way we use it to analyze spin effects, and then 
explicity compute the first next-to-leading interactions for generic
p-branes. We find in general, besides the universal $v^4/r^{7-p}$ term, a spin-orbit 
like coupling whose expansion in velocity gives rise to a long range potential
$v^3/r^{8-p}$, and a spin-spin effect $\sim v^2/r^{9-p}$.
Analogously, higher effects lead to potentials 
of order $v/r^{10-p}$ and $1/r^{11-p}$. 
These leading effects, like in the universal case, are reproduced by
a one-loop computation restricted to the massless sector of the
open strings stretched between the branes.
In particular, for the D0 brane case, this means that a one-loop
(M)atrix theory \cite{bfss} computation in fermionic or more general bosonic
backgrounds, corresponding to the studied relative polarizations, 
should reproduce these effects. 

\section{Supersymmetric D-branes}
In this section we study the spin dependence of D-brane potentials
by using the technique of boundary states in the Green-Schwarz formalism,
following in particular ref.\cite{gregut1}.\\ 
D-branes are solitonic BPS saturated configurations of type IIA(B)
superstring theory. They are arranged in short-multiplets, that in terms of
the little group $SO(9)$ decompose in ${\bf 128+84+44}$, that represent
repectively a massive spin 3/2 fermion
together with a third-rank antisymmetric and a spin 2 bosonic fields.
D-brane interactions, mediated by open strings stretched between them, can be
interpreted in the dual channel due to exchange of closed fields of
which they are sources. Altough many of our results are valid at all
scales, we are mainly interested in this paper to the large distance behaviour
of D-brane interactions, so that our analysis will be performed from the
closed string point of view. D-brane boundary state techniques are indeed
quite useful in this context; a (D)p-brane boundary state $|  B\rangle$
is indeed an object that encodes all the (infinite) couplings between a
(D)p-brane, considered as a classical source, and the closed string
states emitted by it. In the Green-Schwarz formalism, where supersymmetry
is manifest, a boundary state can be defined to be the state preserving the
linear combination of supercharges
\bea Q_{\eta}^a\, |  B,\eta\rangle 
&\equiv& (Q^a+i\eta M_{a\dot{b}}\tilde{Q}^{\dot{b}})\,|B,\eta\rangle
=0 \nonumber \\
Q_{\eta}^{\dot{a}}\, |  B,\eta\rangle 
&\equiv& (Q^{\dot{a}}+i\eta M_{\dot{a}b}\tilde{Q}^{b})\,|B,\eta\rangle
=0\label{bdef}
\eea
valid in type IIA theory for p even; the case of type IIB (p odd) is
easily recovered by switching the dotted and undotted indices in the
right-moving charges $\tilde{Q}$. We borrow in (\ref{bdef}) 
the notation and definitions of \cite{gregut1}, that will
be used throughout all the paper. The solution for $|  B,\eta\rangle$ 
is then given by:
\be \label{boundarystate}
|  B,\eta\rangle =\exp\sum_{n>0}\left(   {1\over n}  M_{ij}
\alpha^i_{-n}\tilde{\alpha}^j_{-n} -
i\eta  M_{a\dot{b}}S^a_{-n}\tilde{S}^{\dot{b}}_{-n}\right)|B_0,\eta\rangle \label{b0}\ee
The indices $i$, $a$, $\dot{a}$ run over the vector and the two spinor
representations of $SO(8)$, $\eta=\pm$ label the brane-antibrane 
nature and the zero mode part is represented by
\be
|B_0,\eta\rangle= 
\left(   M_{ij} |  i\rangle \tilde{|j\rangle} -i\eta
M_{\dot{a}b} |
\dot{a}\rangle\tilde{|b\rangle} \right)
\ee
with the $M$'s given by the $8\times 8$ matrices:
\begin{equation}\label{matrices}
M_{ij} = \pmatrix {- I_{p+1}& 0 \cr
      0 &   I_{7-p}  \cr}, \ \ \ 
M_{\dot{a} b} = i\left(\gamma^1\gamma^2 \dots \gamma^{p+1}\right)_{\dot{a} b}, \ \ \ 
M_{a \dot{b}} = i\left(\gamma^1\gamma^2 \dots \gamma^{p+1}\right)_{a\dot{b}}
\end{equation}

In this formalism $X^{+}=x^++p^+\tau$ and $X^-$ always satisfy Dirichlet
boundary conditions, being fixed by the gauge choice. This means in
particular that all D-branes are actually euclidean-branes and that our
considerations are valid for $-1\leq p\leq 7$ (actually $-1\leq p\leq 6$
for moving branes); moreover,
supersymmetry is manifest, but the unbroken lorentz group is just $SO(8)$.
Since we are also interested to consider the dynamics of moving branes, we
should find a way to define the boundary state for moving branes.
Following \cite{billo}, this can be achieved by performing a boost
transformation to the static boundary state (\ref{boundarystate}).
Since it is not trivial to perform a boost in light-cone gauge,
where the boost operator is a non-linear and complicated object, we use
the following trick to overcome this empasse: we perform an analytic
continuation to an euclidean space and identify the ``time'' with one of
the eight transverse directions, say $X^{p+1}$, and then we realize our boost
along a spatial direction, say $X^{p+2}$, by performing the corresponding
$SO(8)$ rotation with parameter $v$. At the end of the computation we then go
back to Minkowski coordinates by identifying the $p+1^{th}$ direction with i
times the time direction and sending $v\rightarrow iv$.
A boundary state boosted by a transverse velocity $v$ in the $p+2^{th}$
direction is then defined by the boosted matrices:
\begin{eqnarray}\label{boostmatrices}
M_{ij}(v)&\equiv& (\sigma_V(v)M\sigma_V(v)^T)_{ij}\nonumber\\
M_{\dot{a}b}(v)&\equiv& (\sigma_s(v)M\sigma_c(v)^T)_{\dot{a}b} \\
M_{a\dot{b}}(v)&\equiv& (\sigma_c(v)M\sigma_s(v)^T)_{a\dot{b}} \nonumber
\eea
where
\bea\sigma_V(v)&=&\pmatrix {I_{p}& 0&0&0 \cr
      0 &\cos v&-\sin v&0\cr
      0 & \sin v & \cos v & 0\cr
      0 & 0 & 0 & I_{6-p}  \cr}    \nonumber \\
\sigma_s(v)&=&\cos (v/2)\,\delta_{\dot{a}\dot{b}}
-\sin (v/2)\,\gamma^{[p+1 p+2]}_{\dot{a}\dot{b}}, \\
\sigma_c(v)&=&\cos (v/2)\,\delta_{a b}
-\sin (v/2)\,\gamma^{[p+1 p+2]}_{a b} \nonumber \eea
represent the $SO(8)$ rotations on the vector and spinor representations.
   
The universal part of the potential between D-branes moving with
relative velocity $v$ is then easily read from the cylinder
computation \footnote{Since we will always consider in the following
branes of the same charge, we omit the $\eta$ index, that is fixed
to be plus.} \cite{gregut1}:
\begin{equation}
\int_0^\infty dt \
\langle B,x|e^{-2p^+(P^--p^-)t}
| B,y,v \rangle,
\label{cyli}
\end{equation}
where $p^-=i\partial/\partial x^+$ and the boundary states in position space
are given in terms of the momentum states as
\be | B,x \rangle=\int\frac{d^{9-p}q}{(2\pi)^{9-p}}\,e^{iq\cdot x}\,|B,q\rangle
\label{ft} \ee
and
\begin{equation}
P^-=\frac{1}{2p^+}(p^i)^2+\frac{1}{2p^+}\sum_{n=1}^\infty\left(
\alpha^i_{-n} \alpha^i_{n}+\tilde{\alpha}^i_{-n} \tilde{\alpha}^i_{n}  
+ n \,S^a_{-n} S^a_{n}+ n \,\tilde{S}^{\dot{a}}_{-n}\tilde{S}^{\dot{a}}_{-n}
\right) \nonumber \end{equation}
is the Hamiltonian in light-cone gauge. It is a straightforward exercise to
see that eq.(\ref{cyli}), together with the matrices (\ref{boostmatrices}),
reproduce the Bachas formula \cite{bach} after the spin-structures sum. It is
worth while, however, to show how arise the dependence (\ref{vrv}) within
this formalism. Considering only the zero mode part of the boundary state,
we have
\bea &&\langle B_0,v=0|B_0,v \rangle={\rm Tr}_V[M(0)^TM(v)]-{\rm Tr}_S\,
[M(0)^T M(v)]= \nonumber  \\
&=&6+2 \cos 2v-\cos v\,{\rm Tr}\,I + \sin v\,{\rm Tr}\,(\gamma^{p+1}\gamma^{p+2})
=6+2 \cos 2v-8 \cos v  \eea
where the subscript $V,S$ indicate respectively the trace on the vectorial and
spinorial indices.
After analytic continuation, up to a factor two this is just the velocity dependence
of eq.(\ref{vrv}).  
The $1/r^{7-p}$ factor comes from integrating in momentum space and reproduces simply
the scalar massless propagator in the space transverse to the two (D)p-branes.

Besides this universal force, D-branes feel their spin nature through
non-minimal couplings, as seen in the introduction. We construct the currents
$J_M^{spin}$ by applying the broken supercharges $Q^{-}$ to the boundary
state $|B\rangle$. The best way to see that these new boundary states really
encode next-to-leading interactions of D-branes with the bulk fields is
by computing one-point functions of closed vertex operators on a disk with
insertion of supercharges to the boundary. This has been already done for
the $p=-1$ D-instanton in ref.\cite{gregut2}, (eqs.47-50), in the covariant formalism
and for the massless states. They found that, among the usual universal
coupling, the insertion of broken supercharges on the boundary of the disk
allows new couplings with different
closed string states, for which D-branes are in general neutral. In particular, all 
the terms with even numbers of
insertions are formed from powers of the matrix
\be A^{\mu\nu}=\bar{\epsilon}\gamma^{[\mu\nu\rho]}\epsilon\,q_{\rho}
\label{amn} \ee
where $\epsilon$ is the 16-component Majorana-Weyl spinor, parameter of the
supersymmetry, and $q$ is the momentum of the emitted closed string state.
The generalization of this result for $p>-1$ is straighforward; the
presence of Neumann, as well as Dirichlet, boundary conditions will be introduced
(in light-cone gauge) by the $M_{IJ},M_{a\dot{b}}$ matrices that take into
account of the fermionic and bosonic correlators on the disk.

The reformulation in terms of light-cone boundary states 
can also be easily performed.
It simply corresponds to apply a bunch of supercharge pairs, one dotted and one
undotted, to the boundary state (\ref{b0})\footnote{Because of the $q^+$-integration,
a non trivial contribution is obtained only for insertion of dotted-undotted pairs.}.
The usual boundary state $|B\rangle$ represents then the universal leading
couplings of {\it all} the D-branes in the supermultiplet with the bulk
fields, while the other boundary states
$Q^{a_1-}Q^{\dot{a}_1-}|B_0\rangle$, etc.
encode the next-to-leading couplings, different for each state of the
multiplet, i.e. spin effects. If we want to consider brane-potentials
where there is no change in the external states, considered as classical
and heavy, we have to restrict our analysis to products of an even
number of $Q^-$'s applied to the boundary state.
We will study in the next two sections the first next-to-leading 
effects encoded in a cylinder or annulus computation with the
insertions of up to eight supercharges.
For the case of D0-brane potentials, this includes the interactions that in
eleven-dimensional supergravity correspond to the non-minimal couplings
of the gravitino with the four-form field strength.

\section{D-brane dynamics}
In this section we apply the general considerations performed before to 
compute some next-to-leading spin effects. Since we are interested to these 
interactions at large distances, where the closed string channel description is valid,
our analysis will be restricted to the zero-mode part of the boundary state $|B\rangle$.
For the same reason, we will consider the supercharges restricted to the
massless modes; we then apply to the boundary state $|B_0\rangle$ a bunch of
$Q^{a-}$ and $Q^{\dot{a}-}$ that are given by:
\bea Q^{a-}&=&(2q^+)^{1/2}(S_0^a-iM_{a\dot{b}}\tilde{S}_0^{\dot{b}})
\nonumber \\
Q^{\dot{a}-}&=&(2q^+)^{-1/2}q_i\gamma^i_{\dot{a}a}
(S_0^a-iM_{a\dot{b}}\tilde{S}_0^{\dot{b}}) \label{s00} \eea
where, according to the general coniderations of last section, we take for a (D)p-brane
the direction $p+1$ as our ``time''. After having applied 
the supercharges, we can boost
the new boundary state along the direction $p+2$ to obtain the generic
moving current.  The $S_0$ operators are realized as usual by
\be S_0^a|i\rangle=\gamma^i_{a\dot{a}}|\dot{a}\rangle/\sqrt{2}, \ \ \ \
    S_0^a|\dot{a}\rangle=\gamma^i_{a\dot{a}}|i\rangle/\sqrt{2} \label{s0} \ee
and the analogous for the right-moving states.
The boundary state obtained by the insertion of
two of these broken supercharges is then the following:
\be
|B\rangle_{a_1 \dot{a_1}}\equiv Q^{-a_1}Q^{-\dot{a_1}}|B\rangle=
M^{a_1 \dot{a_1}}_{ij}|i\rangle\tilde{|j\rangle}+
iM_{\dot{a}b}^{a_1 \dot{a_1}}|\dot{a}\rangle\tilde{|b\rangle}.\nonumber
\ee
where now the a-dependent $M$ matrices are given by
\bea\label{amatrices}
M^{a_1 \dot{a_1}}_{ij}&\equiv&M_{kj}\,q_l\,
\gamma^{[lki]}_{a_1 \dot{a_1}} \nonumber \\
M^{a_1 \dot{a_1}}_{\dot{a}b}&\equiv&
q_j\,[(\gamma^j\gamma^i)_{\dot{a}_1\dot{a}}(M\gamma^i)_{ba_1}-
(\gamma^j\gamma^iM)_{\dot{a}_1b}\gamma^i_{\dot{a}a_1}] \eea
The boost of these matrices is defined as before through 
eqs.(\ref{boostmatrices}). Sustituing the M's (\ref{matrices}) 
in (\ref{amatrices}), a simple algebra leads to the first spin
correction to D-brane potentials 
\be 
\langle B,v|B,v=0\rangle_{a_1 \dot{a_1}}=
2(\gamma^{[p+1,p+2]}\gamma^i)_{a_1\dot{a}_1}\,q_i\,\sin v(\cos v-1) 
\label{2q} \ee
where $q$ is the momentum transfer between the two D-branes.

We immediately see from eqs.(\ref{cyli}) and (\ref{ft}) that eq.(\ref{2q}) 
produces at large distances a spin-orbit like coupling going like $v^3/r^{8-p}$.

The next-to-leading effect (next power in $q$) comes from the insertion of
four supercharges; in this case the amplitude is given by
\be
_{a_1\dot{a}_1}\langle B,v|B,v=0\rangle_{a_2\dot{a}_2}={\rm Tr}_V
\left(M^{a_1 \dot{a_1}}\sigma_v(v)M^{a_2\dot{a}_2}\sigma_v(v)^T\right)
-{\rm Tr}_S\left(M^{a_1
\dot{a_1}}\sigma_s(v)M^{a_2\dot{a}_2}\sigma_c(v)^T\right)\nonumber
\label{4Q}
\ee
where the trace and the matrix multiplication in both terms are 
over the vectorial and spinorial indices respectively.
It is not difficult to see that for any choice of polarizations
$a_{1,2},\dot{a}_{1,2}$, the static force is zero and the first non-vanishing 
contribution goes as $v^2/r^{9-p}$.
In particular the static force cancels,
due to a compensation between the vectorial and spinorial contributions in 
eq.(\ref{4Q}), corresponding in the NS-R formalism, to exchange of NSNS and RR states,
respectively. 
Note that the non-minimal coupling arising from the
eleven dimensional gravitino-four form field strenght interaction is just
encoded in this amplitude.
There is only an other interaction that can produce effects 
of order $1/r^{9-p}$ and it is the one 
obtained by inserting four supercharges on the same boundary state. 
Again, this leads to a potential of order $v^2/r^{9-p}$.

The six and eight supercharge insertions can be analyzed similarly
although the gamma matrix algebra become more laborious. The ending result is however
simple; as already anticipated in the introduction, they give rise in general
to interactions linear in velocity $\sim v/r^{10-p}$ and to a static force 
$\sim 1/r^{11-p}$ respectively.  
As far as the leading contribution of these higher order effects is concerned,
it is easier to show their general dependence in the open string channel, 
as we will see in next section. 

\section{Open string channel}
It is instructive to see how the leading orders of D-brane 
potentials found before are reproduced by the corresponding
annulus computation in the open string framework. 
The spin potentials are represented, as before, by the insertion
of supersymmetric charges in the partition function of the open strings
stretched between the two moving branes. The boundary conditions 
for these open strings\footnote{We recall that the term `p-brane', as in the
preceeding sections, denotes really a (D)p-instanton, related 
to standard (D)p-branes by a Wick rotation.}  
are given by \cite{bach}
\bea\label{bc}
X^{p+2}+v\,X^{p+1}&=&\partial_{\sigma}(v\,X^{p+2}-X^{p+1})=0 \ \ \ \ {\rm at}
\ \ \sigma=0\nonumber\\ 
X^{p+2}&=&\partial_{\sigma}X^{p+1}=0 \ \ \ \ {\rm at} \ \  \sigma=\pi \eea
in the $p+1^{th}$ (time) and $p+2^{th}$ (brane velocity) directions, while they
satisfy the standard Neumann and Dirichlet conditions 
for the remaining $1,..,p$ and $p+3,...,8$ light-cone directions
respectively.\\
Spin potentials can then be read from $2n$-point functions of fermionic vertex
operators at zero momentum, i.e. supercharges, at one-loop:
\be
{\cal A}^{a_i,\dot{a_i}}_{(2n)}\equiv\int{\cal D}X{\cal D}S\,e^{-(S_0+S_v)}
\prod_{i=1}^n Q^{a_i-} Q^{\dot{a_i}-}\label{cal}
\ee
where $S_0$ is the free worldsheet string action and
\be
S_v\equiv v\oint d\tau\left[ (X^{p+1}\partial_{\sigma} X^{p+2}-
\frac{i}{4}(\bar{S}\rho^1\gamma^{[p+1,p+2]}S) \right]
\nonumber \ee
represents the term that twists the usual
Neumann-Dirichlet boundary conditions
in the $(p+1,p+2)$ plane, according to eqs.(\ref{bc}).
$\rho^1$ is the $2\times 2$ matrix as defined in \cite{gsw} and
the functional integration ${\cal D} X {\cal D} S$ in eq.(\ref{cal}) includes also 
grassmannian integrations over the eight
fermionic zero modes of the untwisted $S_0$ action. 
Expanding in powers of $v$, this leads to a vanishing result unless
the eight fermionic zero modes are soaked up. 
If no supercharges are inserted, then, 
the $v-$twisted partition function is zero up to $v^4$, that is the minimum
power in velocity that soak up all the eight zero modes. The t-modulus
integration leads to the standard $1/b^6$ impact parameter dependence for the
universal phase-shift. 
Each pair of supercharges provides two fermionic and one bosonic 
zero modes producing an additional $b t^2$ insertion in the partition function,
the impact parameter $b$ being the zero mode of $\partial_{\sigma}X$, 
appearing in the dotted supercharge. In this way we have generically
\be
(vt)^m\,(S_0^-\gamma^{[p+1 p+2]}S_0^-)^m
t^{2n}\prod_{i=1}^n b_{k}\gamma^{[ijk]}_{a_i\dot{a_i}}
(S_{0}^-\gamma^{[ij]}S_0^-) 
\label{kin}
\ee
where use has been made of the `Fierz' identity
\be S_0^{a-}S_0^{b-}=\frac{1}{16}(S_0^-\gamma^{[ij]}S_0^-)\gamma^{[ij]}_{ab} \ee
with $2(n+m)=8$, which provide the eight fermionic 
zero modes needed in order to get a 
non-vanishing result and
$n$ the number of dotted-undotted pairs of supercharge insertions. 
We are left then with an additional $(bt)^n$ which  
after the t-modulus integration  
leads to spin effects going like
$v^{(4-n)}/r^{7-p+n}$. 
We should recall, however, that the matching between the two channels
is just in these leading orders, and the complete expression
in terms of the twisted theta functions will differ of course for higher 
orders effects, in exactly the same way  happens for the $v^4/r^{7-p}$ universal
term \cite{dkps}. As already mentioned in the introduction, this suggests 
that for the case of D0 branes, a one loop M(atrix) \cite{bfss}
theory computation will be able to capture these leading large distance supergravity 
spin effects, being all fixed by the massless spectrum. \\ 

{\bf Note added}: Once this work was completed, a revised version of
the paper \cite{harv} 
appeared, whose results partially overlap with those presented here. \\ \\
 
{\bf{Acknowledgements}}

We thank J.P. Derendinger for useful discussions and
the Physics Dept. of the University of ${\rm Neuch\hat{a}tel}$, where some of
this work has been done, for its hospitality. We are particularly grateful
to E. Gava and K. S. Narain for a detailed discussion of the ideas
developed in this paper. Work supported in part by
EEC contract ERBFMRXCT96-0045 (OFES: 950856).

\end{document}